\documentstyle [12pt,amssymb]{article}
\begin {document}
\sloppy

\title {Remark on the additivity conjecture
for the quantum depolarizing channel}

\author {G.G. Amosov}

\maketitle

\abstract {We consider bistochastic quantum channels generated by
unitary representations of the discret group. The proof of the
additivity conjecture for the quantum depolarizing channel $\Phi$
based on the decreasing property of the relative entropy is given.
We show that the additivity conjecture holds for the channel $\Xi
=\Psi \circ \Phi $, where $\Psi $ is the phase damping. }

\vskip 0.5cm

\section {Introduction.}

Let $H,\ dimH=l<+\infty ,$ be a complex Hilbert space. Denote
$\sigma (H),\ Proj(H)$ and $I_H$ the set of all states, i.e.
positive unite-trace operators, the set of all one-dimensional
projections and the identity operator in $H$, respectively. By a
quantum channel $\Psi $ in $H$ we mean a linear map on $\sigma
(H)$ such that the conjugate linear map $\Psi ^*$ defined on the
algebra of all bounded operators $B(H)$ on $H$ is completely
positive and unital (preserving the identity operator). If $\Psi
(\frac {1}{l}I_H)=\frac {1}{l}I_H$, then the channel $\Psi $ is
said to be {\it bistochastic}. The entropy upper bound for the
channel $\Psi $ is defined by the formula
$$
C_1(\Psi )=\sup \limits _{x_j\in \sigma (H),\pi} S(\sum \limits
_{j=1}^r\pi_j\Psi (x_j))-\sum \limits _{j=1}^r\pi_jS(\Psi (x_j)),
$$
where $S(x)=-Trxlogx$ is the von Neumann entropy of $x$
and the supremum is taken over all probability distributions
$\pi =(\pi_j)_{j=1}^r,\ 0\leq \pi_j\leq 1,\ \sum \limits
_{j=1}^r\pi_j=1$.  {\it The additivity conjecture} states that for
any two channels $\Phi $ and $\Psi $
$$
C_1(\Phi \otimes \Psi
)=C_1(\Phi )+ C_1(\Psi ).
$$
If the additivity conjecture holds, one
can easily find the capacity $C(\Psi )$ of the channel $\Psi $ by the
formula $C(\Psi )=\lim \limits _{n\to +\infty } \frac {C_1(\Psi
^{\otimes n})}{n}=C_1(\Psi )$ (see \cite {H98}).  In \cite {H98} the
additivity conjecture is proved for so called c-q and q-c channels.
It follows from the definition of $C_1$ that
$$
C_1(\Psi )\leq S(\frac {1}{l}I_{H})-\inf \limits _{x\in \sigma
(H)} S(\Psi (x)).
$$
The
equality at the last formula was stated in \cite {AHW00} for
bistochastic qubit channels, in \cite {Cor02} for so-called qudit
channels, in \cite {H02} for covariant channels.  Moreover it was
proved in \cite {S03} that for an arbitrary channel the additivity
conjecture for the quantity $C_1$ is equivalent to {\it the
additivity conjecture for the entropy infimum}
$$
\inf \limits _{x\in \sigma
(H_1\otimes H_2)} S(\Phi \otimes \Psi (x))=\inf \limits _{x_1\in
\sigma (H_1)} S(\Phi (x_1))+\inf \limits _{x_2\in \sigma (H_2)}S(\Psi
(x_2)).
$$
Fix a number $p>1$. Then one can define $l_p$-norm of the
channel $\Psi $ by the formula $||\Psi ||_p= (\sup \limits _{x\in
Proj(H)}Tr\Psi (x)^p)^{\frac {1}{p}}$.  It is shown in \cite {AHW00}
that the additivity conjecture is closely connected with the
following {\it multiplicativity conjecture} which states
for two channels $\Phi $ and $\Psi $
$$
||\Phi
\otimes \Psi ||_p= ||\Phi ||_p||\Psi ||_p.
$$
In particular, if the multiplicativity
conjecture holds for $p$ close to $1$, the additivity
conjecture is true also. The multiplicativity conjecture for the
depolarizing channel is proved in \cite {AH02} for integer
numbers of the index $p$. The additivity and multiplicativity
conjectures are shown to be true
for unital qubit channels in \cite {C01},
for the quantum depolarizing channel in \cite {C02} and
for the entanglement-breaking quantum channels in \cite {S02}.
In \cite {HS03} it is considered the connection of the
additivity conjecture for arbitrary channels and
for constrained channels.
Nevertheless, it was no way allowing to check the conjectures
for an arbitrary channel. In \cite {HW02} it is given a counterexample
to the multiplicativity conjecture. So one can expect that
the additivity conjecture can be not true for some channels.

We investigate bistochastic quantum channels generated by
unitary representation of the discret group.
For two states $\rho _1$ and $\rho _2$
let $S(\rho _1,\rho _2)=
Tr\rho _1log\rho _1-Tr\rho _1log\rho _2$ be a relative
entropy.
The decreasing property of the
relative entropy $S(\rho _1,\rho _2)$ states (see \cite {Ohya})
$$
S(\Xi (\rho _1),\Xi (\rho _2))\leq S(\rho _1,\rho _2)
$$
for a quantum channel $\Xi $ and any two states
$\rho _1$ and $\rho _2$.
In particular, if the channel $\Xi $ is bistochastic, it
means that
$$
S(\Xi (\rho ))\geq S(\rho )
$$
for an arbitrary state $\rho $. The quantum depolarizing channel
$\Phi $ is a unique channel satisfying the covariance property
$U\Phi (x)U^*=\Phi (UxU^*)$ for any unitary operator $U$. We
prove the additivity conjecture for $\Phi $ by means of the
decreasing property of the relative entropy. We show that the
additivity property of the entropy upper bound for the
depolarizing channel $\Phi $ allows to prove the additivity
conjecture for the channels $\Xi =\Psi \circ \Phi $, where $\Psi
$ is the phase damping.

\section {The main result.}

{\it The quantum depolarizing channel} is defined by the formula
$\Phi (x)=(1-p)x+\frac {p}{l}I_{H},\ x\in \sigma (H),\ 0<p\leq
\frac {l^{2}}{l^{2}-1}$. Fix the basis $e_j,\ 1\leq j\leq l,$ in
$H$ and define a linear map $\Psi $ by the formula
$$
\Psi (|e_s><e_j|)=q_{|s-j|}|e_s><e_j|,\ 1\leq s,j\leq l,\
|s-j|<l-1,
$$
$$
\Psi (|e_1><e_l|)=q_1|e_1><e_l|,\ \Psi (|e_l><e_1|)=q_1
|e_l><e_1|,
$$
$$
q_0=1,\ 0\leq q_j\leq 1,\ 1\leq j\leq l-1. \eqno (1)
$$
If $\Psi $ is completely positive, then we shall call its
restriction to $\sigma (H)$ by {\it a phase damping}. The phase
damping is a bistochastic channel.

In \cite {C02} it is proved that the additivity and
multiplicativity conjecture hold for the depolarizing channel. We
shall give an alternative proof of this statement. Moreover, we
shall prove the following theorem:

{\bf Theorem.} {\it The additivity conjecture holds for any
channel of the form $\Xi =\Psi \circ \Phi $, where $\Phi $ is the
quantum depolarizing channel and $\Psi $ is the phase damping.}

Denote $\Xi =\Xi (\lambda _1,\lambda _2,\lambda _3)$ the
bistochastic qubit channel with the parameters $(\lambda
_1,\lambda _2,\lambda _3)$ acting by the formula
$$
\Xi \left \{\left (\begin {array}{cc}a&b+ic\\b-ic&d\end {array}
\right )\right \}=\left (\begin {array}{cc}\frac {1+\lambda
_3}{2} a+\frac {1-\lambda _3}{2}d&\lambda _1b+i\lambda _2c\\
\lambda _1b-i\lambda _2c&\frac {1-\lambda _3}{2}a+ \frac
{1+\lambda _3}{2}d\end {array}\right ),
$$
where matrices written in some basis $e_1,e_2$ of the Hilbert
space $H,\ dimH=2,\ a,b,c,d\in {\mathbb R}$. Let us assume that
$\lambda _1=\lambda _2$ and $|\lambda _{1}|\leq \lambda _3$. Then
$\Xi $ can be represented as $\Xi (\lambda _1,\lambda _1,\lambda
_3)= \Xi (\frac {\lambda _1}{\lambda _3},\frac {\lambda
_1}{\lambda _3},1) \circ \Xi (\lambda _3,\lambda _3,\lambda _3)$.
Notice that $\Xi (\lambda _3,\lambda _3,\lambda _3)$ is the
depolarizing channel and $\Xi (\frac {\lambda _1}{\lambda
_3},\frac {\lambda _1} {\lambda _3},1)$ is the phase damping. It
follows from Theorem that the additivity and multiplicativity
conjectures hold for the qubit channels of the form defined above.

\section {Bistochastic quantum channels generated by unitary
representations of the discret group.}

Let $G$ and $G_0\subset G$ be a finite discret group and its
normal subgroup. Denote by $e$ the unit of the group $G$. Suppose
that $|G|=l^2,\ |G_0|=l$, and there exists an irreducible
projective unitary representation $g\to U_g$ of the group $G$ in
the Hilbert space $H$. Consider the factor-group $G/G_0$
consisting of equivalency classes $[g]=\{gg_0\ |\ g_0\in G_0\}$.
It follows from the definition that $|G/G_0|=l$. Take one element
in each class $[g]$ and enumerate these elements in an arbitrary
order. Then we get the set $G_1=\{g_1\equiv e,g_2,\dots ,g_l\}$
such that the classes $[g_s]$ generate $G/G_0$. We shall suppose
that $G_1$ is a subgroup of $G$. Notice that there exist $l$ ways
to construct the group $G_{1}$ accordingly to a number of elements
in each coset $[g]$.

{\bf Example.} The discret Weyl group.

Put $G={\mathbb Z}_l \oplus {\mathbb Z}_l$ and $G_0={\mathbb
Z}_l\oplus 0,\ G_1=0\oplus {\mathbb Z}_l$. Notice that $G_0$ and
$G_1$ are normal subgroups of $G$. On the other hand, every
element $1\oplus s$ generates a subgroup of $G$ for $0\leq s\leq
l-1$. We shall denote this subgroup by $G_{0k},\ 0\leq k\leq l-1,$
such that $G_{00}\equiv G_0$. Thus, $G_{0k}=\{s\oplus sk,\ 0\leq
s\leq l-1\}$. Notice that if $0\leq k\leq l-1$ and $s\neq 0$ the
elements $s\oplus sk\in G_{0k}$ run the coset $[s\oplus 0]$ of
the factor-group $G/G_{1}$. Fix a basis $(e_j)_{j=1}^{l}$ in the
Hilbert space $H$ and define unitary operators $U_g,\ g\in G,$ as
follows
$$
U_ge_k=e_{(k+s)mod(l))},\ g=s\oplus 0\in G_0,
$$
$$
U_ge_k=e^{\frac {2\pi isk}{l}}e_k,\ g=0\oplus s\in G_1.
$$
Then, the formula $U_{k\oplus s}=U_{k\oplus 0}U_{0\oplus s},\
k\oplus s\in G$, defines an irreducible projective unitary
representation of the group $G$. $\Box $

Let us define completely positive linear maps
$E_0$ and $E_1$ as follows
$$
E_k(x)=\frac {1}{l}\sum \limits _{g\in G_k}^lU_{g}xU_{g}^*,\ x\in
\sigma (H),\ k=0,1. \eqno (2)
$$
The map $E_k$ is a conditional expectation to the
algebra of fixed elements ${\cal A}_k=\{x\in B(H)\ |\
U_{g}xU_{g}^*=x,\ g\in G_k\}$, $k=0,1$.  Because the representation
$g\to U_g$ is irreducible, we get the equality
$$
\sum \limits _{k=1}^lU_{g_k}E_0(x)U_{g_k}^*=I_H,\ x\in \sigma (H).
\eqno (3)
$$
Hence, the operators $X_k=U_{g_k}E_0(x)U_{g_k},\ 1\leq k\leq l,$
form a (non orthogonal, in general) resolution of the
identity.
Take a probability distribution $\mu =\{\mu _g,\ g\in G\}$
on the group $G$ and
define a bistochastic channel $\Phi $ by the formula
$$
\Phi (x)=\sum \limits _{g\in G}\mu _gU_gxU_g^*,\ x\in \sigma (H).
\eqno (4)
$$
Let us define a probability distribution $\lambda =\{\lambda _k,\
1\leq k\leq l\}$ on the factor-group $G/G_0$ as follows
$$
\lambda _k=\sum \limits _{g\in G_0}\mu _{g_kg_0},\ 1\leq k\leq l.
$$
Consider the channel $\Phi \otimes Id$ in the
Hilbert space $\tilde H=H\otimes K$. Here and in the
following we denote by $Id$ the ideal (identity) channel.

{\bf Proposition 1.} {\it Suppose that the numbers $\frac {\mu
_{g_kg}} {\lambda _k}=\epsilon _g,\ g\in G_0,$ do not depend on a
choice of $k,\ 1\leq k\leq l$. Then, the following inequality
holds,
$$
S((\Phi \otimes Id)(x))\geq S(\sum \limits _{k=1}^l\lambda _k
(U_{g_k}\otimes I_{K})x(U_{g_k}^*\otimes I_{K})),\ x\in \sigma
(H\otimes K).
$$
}

Proof.

Let us define a bistochastic channel $\Psi $ by the
formula
$$
\Psi (x)=\sum \limits _{g\in G_0}\epsilon _gU_gxU_g^*,\
x\in \sigma (H).
$$
Then,
$$
S(\Phi \otimes Id(x))=S((\Psi \otimes Id)(
\sum \limits _{k=1}^n\lambda _kU_{g_k}xU_{g_k}^*))\geq
S(\sum \limits _{k=1}^n\lambda _kU_{g_k}xU_{g_k}^*)
$$
by means of the decreasing property of the relative entropy.
$\Box $

Now put $\Phi (x)=\sum \limits _{k=1}^l\lambda
_kU_{g_k}xU_{g_k},\ x\in \sigma (H)$, $A_1=\{x\ |\
U_{g_k}xU_{g_k}^*=x,\ 1\leq k\leq l\}$ and $E_1$ is the conditional
expectation $(1)$ to $A_1$.  Then,
$\tilde E_1(x)=E_1\otimes Id(x)=\frac
{1}{l}\sum \limits _{g\in G_0}^l
(U_{g}\otimes I_K)x(U_{g}\otimes
I_K),\ x\in \sigma (H),$ is a conditional expectation to the algebra
$A_1\otimes B(K)$. Because $A_1$ is the algebra of fixed elements
for the action of the group $(U_{g_k})_{k=1}^l$, we get
$$
E_1\circ \Phi (x)=\Phi \circ E_1(x)=E_1(x),\ x\in \sigma (H).
\eqno (5)
$$
Pick up the orthogonal resolution of the identity
$(P_k)_{k=1}^l$ generating $A_1$.

{\bf Proposition 2.}{\it
$$
S((\Phi \otimes Id)(x))\geq -\sum \limits _{k=1}^l\lambda _k
log\lambda _k
$$
$$
-\sum \limits _{k=1}^l Tr_H((P_k\otimes I_{K})\tilde
E_1(x))logTr_H((P_k\otimes I_{K}) \tilde E_1(x))-log(l).
$$
}

Proof.

Let us define a quantum channel $\Xi _x$ in the
Hilbert space $H\otimes K$ by the formula
$$
\Xi _x(\rho )=\sum \limits _{k=1}^l Tr((P_k\otimes I_K)\rho )
(U_{g_k}\otimes I_K)x(U_{g_k}^{*}\otimes I_K),\ \rho \in \sigma
(H\otimes K).
$$
Put $\rho =\sum \limits _{k=1}^l\lambda _kP_k\otimes Q$,
$\overline \rho =\frac {1}{l}\sum \limits _{k=1}^lP_k\otimes Q=
(\frac {1}{l})I_H\otimes Q$,
where $Q\in Proj(K)$. Then,
$\Xi _x(\rho )=(\Phi \otimes Id)(x)$ and
the relative entropy
$$
S((\Phi\otimes Id)(x),\tilde E_1(x))=
S(\Xi _x(\rho ),\Xi _x(\overline \rho ))\leq
$$
$$
S(\rho ,\overline \rho )=Tr(\rho log\rho )- Tr(\rho log \overline
\rho )=\sum \limits _{k=1}^l\lambda _k log\lambda _k+log(l).
\eqno (6)
$$
On the other hand,
$$
S((\Phi \otimes Id)(x),\tilde E_1(x))=\sum \limits _{k=1}^l
Tr((\Phi \otimes Id)(x)log(\Phi \otimes Id)(x))-
$$
$$
Tr((\Phi \otimes Id)(x)log\tilde E_1(x))=
-S((\Phi \otimes Id)(x))
$$
$$
-Tr(\tilde E_1((\Phi \otimes Id)(x))log\tilde E_1(x))=
$$
$$
-S((\Phi \otimes Id)(x))-Tr\tilde E_1(x)log\tilde E_1(x). \eqno
(7)
$$
Here we use the equality $\tilde E_1((\Phi \otimes Id)
(x))=\tilde E_1(x)$ in virtue of $(5)$. Notice that $\tilde
E_1(x)=\sum \limits _{k=1}^lP_k\otimes x_k$, where $x_k\in \sigma
(K)$. Then, $P_k\otimes x_k=\tilde E_1(x)P_k\otimes I_K$. It
follows that
$$
x_k=Tr_H((P_k\otimes I_K)\tilde E_1(x)),\
1\leq k\leq l.
$$
In this way,
$$
\tilde E_1(x)=\sum \limits _{k=1}^lP_k\otimes Tr_H((P_k\otimes
I_K)\tilde E_1(x)),\ x\in \sigma (H\otimes K). \eqno (8)
$$
To complete the proof, it is sufficiently to substitute $(8)$ to
$(7)$ and compare the result with $(6)$. $\Box $

\section {The quantum depolarizing channel.}

Consider the quantum depolarizing channel $\Phi (x)=(1-p)x+\frac
{p}{l}Tr(x),\ x\in \sigma (H)$. Due to the property $\frac
{1}{l}\sum \limits _{g\in G}U_gxU_g^*=I_H$ the channel $\Phi $
can be represented as $\Phi (x)=(1-\frac {l^2-1}{l^2}p)x+\sum
\limits _{g\in G,g\neq e} \frac {p}{l^2}U_gxU_g$. Hence, it has
the form $(3)$ with $\mu _e=1-\frac {l^2-1}{l^2}p,\ \mu _g=\frac
{p}{l^2},\ g\neq e$. Let $G$ be the discret Weyl group. Any
element $g\neq e$ of the group $G$ belongs to one and only one
subgroup $G_{0k}$ or $G_{1}$. On the other hand, two elements
$g_{k}\in G_{0k}$ and $g_{r}\in G_{0r}$ are connected by certain
element $g\in G_{1}$ such that $g_{r}=g_{k}+g$. Thus, the direct
sum $\oplus _{0\leq k\leq l-1}G_{0k}$ consists of $l$ units and
one by one non-unit elements of $G$ except elements of $G_{1}$.
The direct sum $\oplus _{g\in G_{1},g\neq e,0\leq k\leq
l-1}\{g+G_{0k}\}$ consists of $l$th copies of non-unit elements
of $G_{1}$ and $(l-1)$th copies of non-unit elements of
$\{G_{0k},\ 0\leq k\leq l-1\}$. Put $c_0=\frac {1}{l}\frac
{1-\frac {l^2-1}{l^2}p}{1-\frac {l-1}{l}p},\ c_1=\frac
{1}{l}\frac {\frac {p}{l^2}} {1-\frac {l-1}{l}p}$, $\lambda
_0=1-\frac {l-1}{l}p,\ \lambda _k= \frac {p}{l},\ 1\leq k\leq
l-1$.  Then, the depolarizing channel $\Phi $ can be represented
as
$$
\Phi (x)=c_0\sum \limits _{k=0}^{l-1}\Phi _k(x)+c_1\sum \limits
_{k=0}^{l-1} \sum \limits _{s=1}^{l-1}U_{0\oplus s}\Phi
_k(x)U_{0\oplus s}^*, \eqno (9)
$$
where
$$
\Phi_k (x)=\sum \limits _{s=0}^{l-1} \lambda _sU_{s\oplus
sk}xU_{s\oplus sk}^*,\ 0\leq k\leq l-1.
$$
Notice that the first sum in (9) includes the actions of the
elements from the direct sum $\oplus _{0\leq k\leq l-1}G_{0k}$,
while the second sum includes the actions of the elements from
$\oplus _{g\in G_{1},g\neq e,0\leq k\leq l-1}\{g+G_{0k}\}$.

{\bf Proposition 3.} {\it Let $\Phi $ be the quantum depolarizing
channel. Then, given a state $x\in \sigma (H\otimes K)$
there exists
the projection $P\in Proj(H)$ such that
$\rho =lTr_H((P\otimes Id)x)\in \sigma (K)$ and
$$
S(\Phi \otimes Id(x))\geq -(1-\frac {l-1}{l}p)log(1- \frac
{l-1}{l}p)-(l-1)\frac {p}{l}log\frac {p}{l}+ S(\rho ).
$$
}

Proof.

Pick up a unitary operator $W$ in the Hilbert space $H$
such that the projection $y=(W\otimes I_K)x(W^*\otimes I_K))$
has the form
$$
y=\frac {1}{l}\sum \limits _{k=1}^l P_k\otimes x_k,\ x_k\in
\sigma (K). \eqno (10)
$$
This property is equivalent to the condition $WTr_K(x)W^*=\frac
{1}{l}I_H$. Using the covariance of the depolarizing channel we
obtain $(\Phi \otimes Id)(x)=(W^*\otimes I_K)(\Phi \otimes Id)
(y)(W\otimes I_K)$. The representation $(9)$ gives us the estimate
$$
S((\Phi \otimes Id)(x))=
S((\Phi \otimes Id)(y))\geq \inf \limits _{0\leq k\leq l-1}
S((\Phi _k\otimes Id)(y)).
$$
Applying Proposition 2 to the projection
$y$ we get the inequality
$$
S((\Phi _k\otimes Id)(y))\geq -(1-\frac {l-1}{l}p)log(1- \frac
{l-1}{l}p)-(l-1)\frac {p}{l}log\frac {p}{l}
$$
$$
-\sum \limits _{k=1}^l Tr_H((P_k\otimes I_K)\tilde E_1(y))
logTr_H((P_k\otimes I_K)\tilde E_1(y))- log(l). \eqno (11)
$$
The condition $(10)$ results in the formula $Tr_H((P_k\otimes
I_K)\tilde E_1(y))=\frac {1}{l}y_k,\ y_k\in \sigma (K)$. Choose
$k_0$ such that $\inf \limits _{1\leq k\leq l}S(y_k)=S(y_{k_0})$.
Put $P=W^*P_{k_0}W$, then $(11)$ implies that
$$
S((\Phi _k\otimes I_K)(x))\geq
-(1-\frac {l-1}{l})log(1-
\frac {l-1}{l}p)-(l-1)\frac {p}{l}log\frac {p}{l}
$$
$$
+S(lTr_H(P\otimes I_K)x).
$$
The result follows.
$\Box $

{\bf Proposition 4.} {\it
Let the following inequalities hold,
$$
q_j\leq \frac {1}{l-1}(1+\sum \limits _{j=1}^{l-2}q_j)\equiv q,
\ 1\leq j\leq l-2,
$$
then the map $\Psi $ defined by means of
$(1)$ is completely positive.
}

{\bf Remark.} If
$q_j=Q=const,\ 1\leq j\leq l-2,$ then $q=\frac {1}{l-1}
(1+(l-2)Q)\geq Q$ and the condition of Proposition 4 is
satisfied.

Proof.

The map $\Psi $ can be represented as
$$
\Psi(x)=qx+\sum \limits _{s=1}^{l-2}
\sum \limits _{r,j:|r-j|=s,r<j}(q-q_s)(|e_r><e_r|-|e_j><e_j|)x\cdot
$$
$$
\cdot (|e_r><e_r|-|e_j><e_j|)+
$$
$$
(q-q_1)(|e_1><e_1|-|e_l><e_l|)x(|e_1><e_1|-|e_l><e_l|).
\eqno (12)
$$
It follows from $(12)$ that
the property $q_j\leq q$ garantees the complete
positivity for $\Psi $. $\Box $

Proof of Theorem.

The additivity conjecture for the depolarizing channel
immediately follows from Proposition 3.
The decreasing property of the entropy implies that
$$
S(\Xi ^{\otimes n}(x))\geq S(\Phi ^{\otimes n}(x)).
\eqno (13)
$$
Hence, to derive the result we need to prove equalities in
$(13)$. It follows from the definition $(1)$ of the phase damping
that $\Psi $ maps the projections $Q_j=|e_j><e_j|$ to itself,
$\Psi (|e_j><e_j|)=|e_j><e_j|$. So $S(\Psi \circ \Phi
(Q_j))=S(\Phi (Q_j))=\inf \limits _{P\in Proj(H)} S(\Phi (P))$.
The result follows. $\Box $

{\bf Acknowlegments.} The author is grateful to Professor
A.S. Holevo for a careful reading of the text and
many fruitful
discussions. The work is partially supported
by INTAS 00-738.

\end {document}